\documentclass[usenatbib]{mnras}

\usepackage{caption}
\usepackage{natbib}
\usepackage{float}
\usepackage[T1]{fontenc}
\usepackage{aecompl}
\usepackage{amsmath,amsfonts,amssymb}
\usepackage[pdftex]{graphicx}
\usepackage{aas_macros}
\usepackage[usenames]{xcolor}
\usepackage{color}
\usepackage{textcomp,gensymb}
\usepackage{times}
\usepackage{subfig}

\title[Dusty clumps in circumbinary discs]{Dusty clumps in circumbinary discs}

\author[P.P. Poblete, N. Cuello \& J. Cuadra]{
Pedro P. Poblete$^{1,2}$,
Nicol\'as Cuello$^{1,2}$ and
Jorge Cuadra$^{1,2}$
\\
$^{1}$Instituto de Astrof\'isica, Pontificia Universidad Cat\'olica de Chile, Santiago, Chile,\\
$^{2}$N\'ucleo Milenio de Formaci\'on Planetaria (NPF), Chile.}

\begin{document}
\date{Accepted 2019 August 15. Received 2019 August 14; in original form 2019 April 25}

\pagerange{\pageref{firstpage}--\pageref{lastpage}} \pubyear{2019}

\maketitle

\label{firstpage}

\begin{abstract} 
Recent observations have revealed that protoplanetary discs often exhibit cavities and azimuthal asymmetries such as dust traps and clumps. The presence of a stellar binary system in the inner disc regions has been proposed to explain the formation of these structures. Here, we study the dust and gas dynamics in circumbinary discs around eccentric and inclined binaries. This is done through two-fluid simulations of circumbinary discs, considering different values of the binary eccentricity and inclination. We find that two kinds of dust structures can form in the disc: a single horseshoe-shaped clump, on top of a similar gaseous over-density; or numerous clumps, distributed along the inner disc rim. The latter features form through the complex interplay between the dust particles and the gaseous spirals caused by the binary. All these clumps survive between one and several tens of orbital periods at the feature location. We show that their evolution strongly depends on the gas--dust coupling and the binary parameters. Interestingly, these asymmetric features could in principle be used to infer or constrain the orbital parameters of a stellar companion --- potentially unseen --- inside the inner disc cavity. Finally, we apply our findings to the disc around AB~Aurigae.
\end{abstract}



\section{Introduction}

In the last years, the field of planet formation has experienced an unprecedented development thanks to last-generation telescopes. In particular, by combining multi-wavelength observations of protoplanetary discs, it has been possible to map the dust distribution for a wide range of grain sizes around young stars. Extreme adaptive optics instruments in large optical/NIR telescopes and radio antennas observing at mm and submm wavelengths played a key role in achieving this task. This shed some light on the very first stages of planet formation within these systems \citep{Avenhaus18,Pinilla18}. Among the now overwhelming number of {\sc alma} observations of protoplanetary discs, the continuum emission detected around HL~Tau is one of the most spectacular \citep{ALMA}. Especially, the numerous gaps observed suggest that planets might have already formed in this young disc \citep{DiPierro15}. Besides the routinely detected gaps, there are also numerous observations of enigmatic structures such as rings, spirals, warps, clumps, and vortices. For instance, the recent {\sc dsharp} survey by \cite{Andrews18} mapped twenty nearby protoplanetary discs at an astonishing resolution of roughly 5~au. However, the rich structure of these systems remains only partly understood from the theoretical point of view \citep{Armitage18}.

Interestingly, a fraction of these circumstellar discs orbit a binary stellar system instead of a single star. These constitute a special category of protoplanetary discs called \textit{circumbinary}. Considering the stellar context, roughly half of the solar-type stars are singles, whereas about 33\% of them form double systems \citep{Raghavan10, Tokovinin14}. Therefore, about a third of the young stellar systems could potentially harbour circumbinary discs (CBDs), along with circumstellar ones. Hence, a proper understanding of circumbinary disc dynamics is of crucial importance \citep{Nixon+2013,Dunhill15}.

The case of the disc around HD~142527 is particularly enlightening in this regard. \cite{Fukagawa06} first detected a disc with several spiral arms and a large inner cavity of roughly 90~au. This disc was initially thought to be orbiting a single star. However, a companion was later discovered inside the inner cavity by \citet{Biller12}. The stellar masses in HD~142527 are 1.8~$M_\odot$ \citep{Gaia2016b} and 0.4~$M_\odot$ \citep{Christiaens18}, so it is an \textit{unequal-mass binary}. Further studies focused on the companion's orbital motion \citep{Lacour16,Claudi19} indicating that the binary is eccentric and likely inclined with respect with the disc. Based on these constraints, \citet{Price18} presented a consistent hydrodynamical model of HD~142527 where the CBD is periodically perturbed by the inner binary. Remarkably, the resulting gaseous and dust structures are in excellent agreement with all the available multi-wavelength observations: i) the spirals and their location \citep{Avenhaus14,Christiaens14}, ii) the cavity size \citep{Perez15}, iii) the dusty clumps along a horseshoe \citep{Casassus15,Boehler17}, iv) the gaseous filaments crossing the cavity \citep{Casassus13}, and v) the shadows \citep{Avenhaus14} --- likely caused by the presence of a misaligned inner disc \citep{Marino15}.

This circumbinary scenario could very well apply to other discs with large inner cavities  exhibiting various asymmetries. For instance, \cite{Ragusa17} explored how unequal-mass (circular) binaries in a coplanar configuration are able to generate lopsided features and horseshoes at the edge of the cavity --- comparable to the ones observed. Alternatively, the presence of vortices has been widely proposed to explain the same asymmetries in protoplanetary discs \citep{Meheut12,Lyra13,Ataiee13,Marel16}. It is worth noting that, in the binary scenario, no vortex is required whatsoever. Regardless of their origin, these azimuthal pressure maxima are expected to efficiently trap dust in the inner disc regions \citep{Birnstiel13}.

The aim of this work is to study the effect of an inclined and eccentric inner binary on the dust content of the surrounding CBD. To do so, we consider relatively high eccentricities ($e_{\rm B}=0.5$ and $e_{\rm B}=0.75$) and different inclinations for the binary, from prograde ($i_{\rm B}=0\degree$) to retrograde ($i_{\rm B}=180\degree$) configurations. The numerical method and the initial setup of our three-dimensional hydrodynamical simulations are described in Section~\ref{Method}. We report our results in Section~\ref{Results}. In Section~\ref{Discussion}, we discuss the formation of dusty clumps, their evolution and how these can be used to infer the presence of a potentially unseen inner companion. Finally, we draw our conclusions in Section~\ref{Conclusion}.

\section{Numerical method}\label{Method}

We perform 3D hydrodynamics simulations of circumbinary discs (CBDs) using the {\sc phantom} smoothed particle hydrodynamics (SPH) code \citep{PricePH18b}. We use the two-fluid method in order to model the interaction between gas and dust particles as described in \citet{Laibe12a,Laibe12b}. Although each fluid is treated independently, gas and dust particles interact with each other through aerodynamical drag forces. This means that the back-reaction from the dust on the gas is included in our calculations.

\subsection{Binary setup}

We consider a binary system where both stars are treated as sink particles \citep{Bate95}. We explore a range of orbital parameters similar to those observed in HD~142527. In particular, we test different combinations of binary inclination ($i_{\rm B}$) and eccentricity ($e_{\rm B}$). The mass ratio between the primary and the secondary stars is fixed at $q=M_1/M_2=0.25$, with $M_1 =2\,M_\odot$ and $M_2=0.5\,M_\odot$. The semi-major axis is set to 40 au (as in \citet{Price18} for HD~142527B). The free parameters in our simulations are $e_{\rm B}$ and $i_{\rm B}$. In this work, we consider the following sets of values: $e_{\rm B}=\{0.50,0.75\}$ and $i_{\rm B}=\{0\degr,30\degr,60\degr,90\degr,120\degr,150\degr,180\degr\}$.  We model each system for a hundred binary orbits. During this evolutionary time the binary orbit does not change significantly, as expected. Therefore, the parameters aforementioned can be considered as constant throughout each simulation. The effect of the different combinations of orbital parameters on the CBD structure will be discussed in more detail in Section \ref{sims}.

\subsection{Disc setup}

\subsubsection{Initial conditions}
\label{sec:initialconditions}

We follow \citet{Price18} and consider a disc setup consistent with the observations of HD~142527. We model the circumbinary disc with $10^6$ gas particles and $10^5$ dust particles, assuming a total gas mass of $0.01\,M_\odot$ and a dust-to-gas ratio of $0.01$. The spatial distribution of both fluids is initially the same. We set the inner and the outer edges at $R_{\rm in} = 90$ au and $R_{\rm out} = 350$ au, respectively. The surface density profile is given by a power law, $\Sigma \propto R^{-1}$. The temperature profile  follows a shallower power law, $T\propto R^{-0.3}$, as suggested by \citet{Casassus15k} for this system. This corresponds to $H/R = 0.06$ at $R_{\rm in}$ and $H/R = 0.1$ at $R_{\rm out}$. We set the SPH viscosity parameter $\alpha_{\rm AV} \approx 0.3$, which gives a mean Shakura--Sunyaev disc viscosity of $\alpha_{\rm SS}\approx0.005$ \citep{Shakura73} (see Sect.~\ref{A3}). We do not consider self-gravity effects nor magnetic fields.

\subsubsection{Dust modelling}\label{sec:dust_pres}

\begin{figure}
\begin{center}
\includegraphics[width=0.5\textwidth]{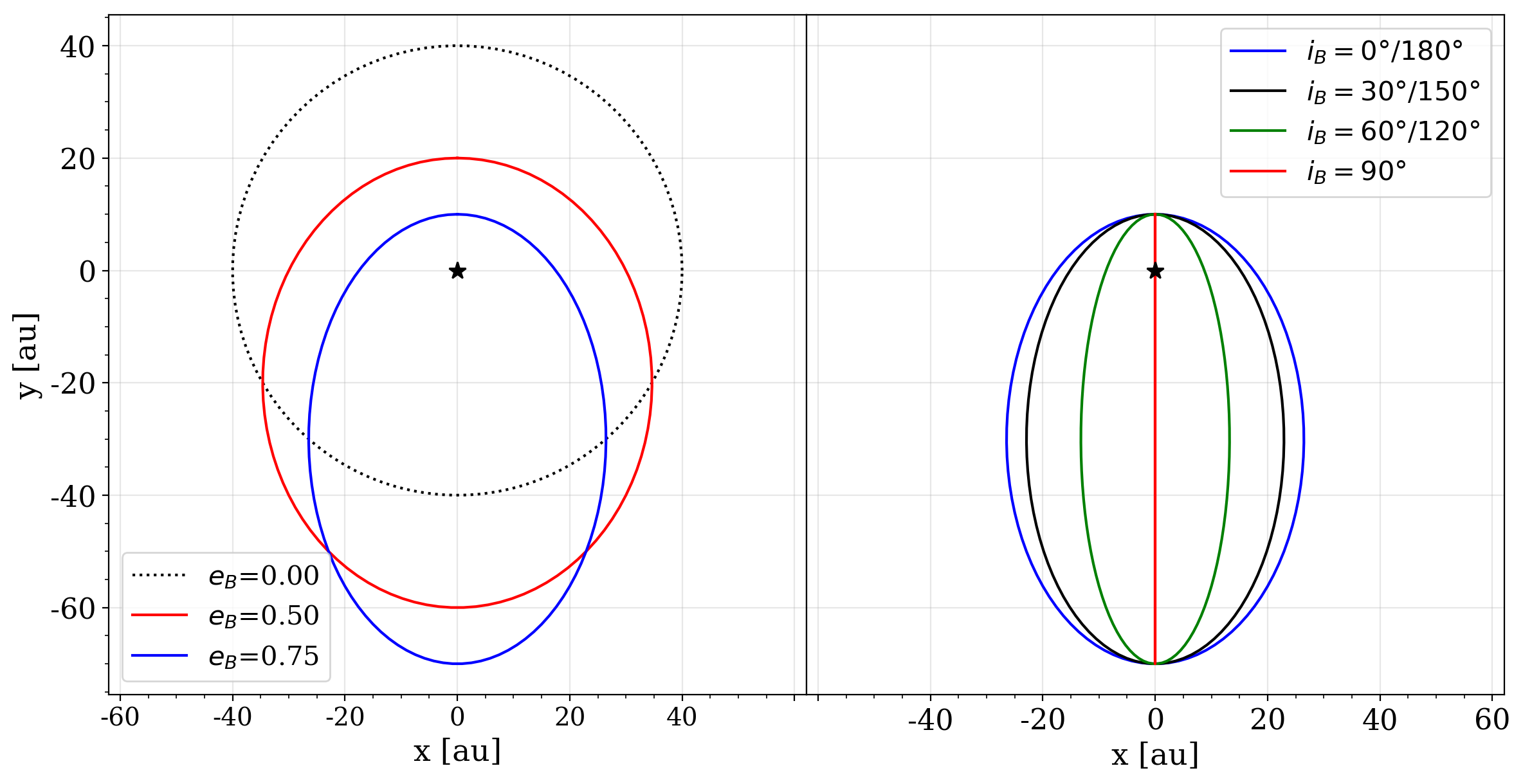}
\caption{Parameter space of orbits modelled. The left panel shows face-on views of the two simulated eccentricities, $e_{\rm B}=0.50$ in red and $e_{\rm B}=0.75$ in blue.  A circular orbit is shown with a dotted line for comparison. The right panel shows the projection of the orbit with  $e_{\rm B}=0.75$ for all eight modelled inclinations.}\label{orbits}
\end{center}
\end{figure}

\begin{table}
\centering
\begin{tabular}{ccc}
\hline
\hline
Orbit name& $e_{\rm B}$ & $i_{\rm B}$\\
\hline
 $e$50-$i$0 &0.50& $0\degree$ \\
 $e$50-$i$30 &0.50& $30\degree$\\
 $e$50-$i$60 &0.50& $60\degree$ \\
 $e$50-$i$90 &0.50& $90\degree$\\
 $e$50-$i$120 &0.50& $120\degree$ \\
 $e$50-$i$150 &0.50& $150\degree$\\
 $e$50-$i$180 &0.50& $180\degree$\\
 \hline
 $e$75-$i$0 &0.75& $0\degree$ \\
 $e$75-$i$30 &0.75& $30\degree$\\
 $e$75-$i$60 &0.75& $60\degree$ \\
 $e$75-$i$90 &0.75& $90\degree$\\
 $e$75-$i$120 &0.75& $120\degree$ \\
 $e$75-$i$150 &0.75& $150\degree$\\
 $e$75-$i$180 &0.75& $180\degree$\\
\hline
\end{tabular}
\caption{List of orbital parameters for each simulation. The parameters are the orbit name, eccentricity $e_{\rm B}$, and inclination with respect to the disc $i_{\rm B}$. }\label{params}
\end{table}

The dust coupling to the gas is well described by the dimensionless quantity called Stokes number (noted $\rm St$). This quantity is defined as the ratio of the orbital time-scale to the drag stopping time\footnote{the time-scale for the drag to damp the local differential velocity between the gas and dust.}. In particular, when ${\rm St} \ll 1$ (${\rm St} \gg 1$) then dust particles are strongly (weakly) coupled to the gas. Interestingly, if $\rm St\sim1$, then the particles are marginally coupled to the gas and experience the fastest radial drift \citep{Weidenschilling77,Nakagawa86}. In all our simulations, the mean free path of the dust particles is greater than their size. Hence, the drag force falls in the Epstein regime \citep{Epstein24} and the Stokes number is given by
\begin{equation}
\label{eq:Stokes}
\mathrm{St} =  \frac{\rho_{\rm s}s}{\rho c_{\rm s}f}\sqrt{\frac{\pi \gamma}{8}} \Omega_{\rm k} ,
\end{equation}
where $\Omega_{\rm k}$ is the Keplerian angular velocity, $\rho_{\rm s}$ is the dust grain density \footnote{we adopt an intrinsic dust grain density equal to 3 $\rm gr\,cm^{-3}$, a typical value used for astrophysical silicates.}, $c_{\rm s}$ the sound speed, $s$ the grain size, $f$ a correction factor for supersonic drag, and $\rho$ the total density $\rho_{\rm g} + \rho_{\rm d}$ (i.e. the sum of the gas and dust volume densities). This prescription is similar to the treatment made by \citet{DiPierro15,DiPierro16} and \citet{Price18}. Here, for computational convenience, we drop the $\rho_{\rm d}$ term in the sum and use instead $\rho = \rho_{\rm g}$ in the code. By doing so, the stopping time is overestimated by a factor of $(1+\epsilon)$, where $\epsilon$ is the dust-to-gas ratio. The value of $\epsilon$ remains always below unity in all our simulations. Therefore, despite using this approximation, we obtain meaningful results for the dust evolution in the disc (at least for the short evolutionary times considered). The dust behaviour as a function of the Stokes number is discussed in Section~\ref{Clumps}.

We chose the grain size for which the particles have a Stokes number close to unity. For the parameters considered in Sect.~\ref{sec:initialconditions}, this size corresponds to $s=1$~mm. The reason for that choice, is that we wish to study the particles that: i) feel the strongest radial drift, and ii) concentrate the most efficiently in the pressure maxima of the CBD.

\begin{figure*}
\begin{center}
\includegraphics[width=.55\textwidth]{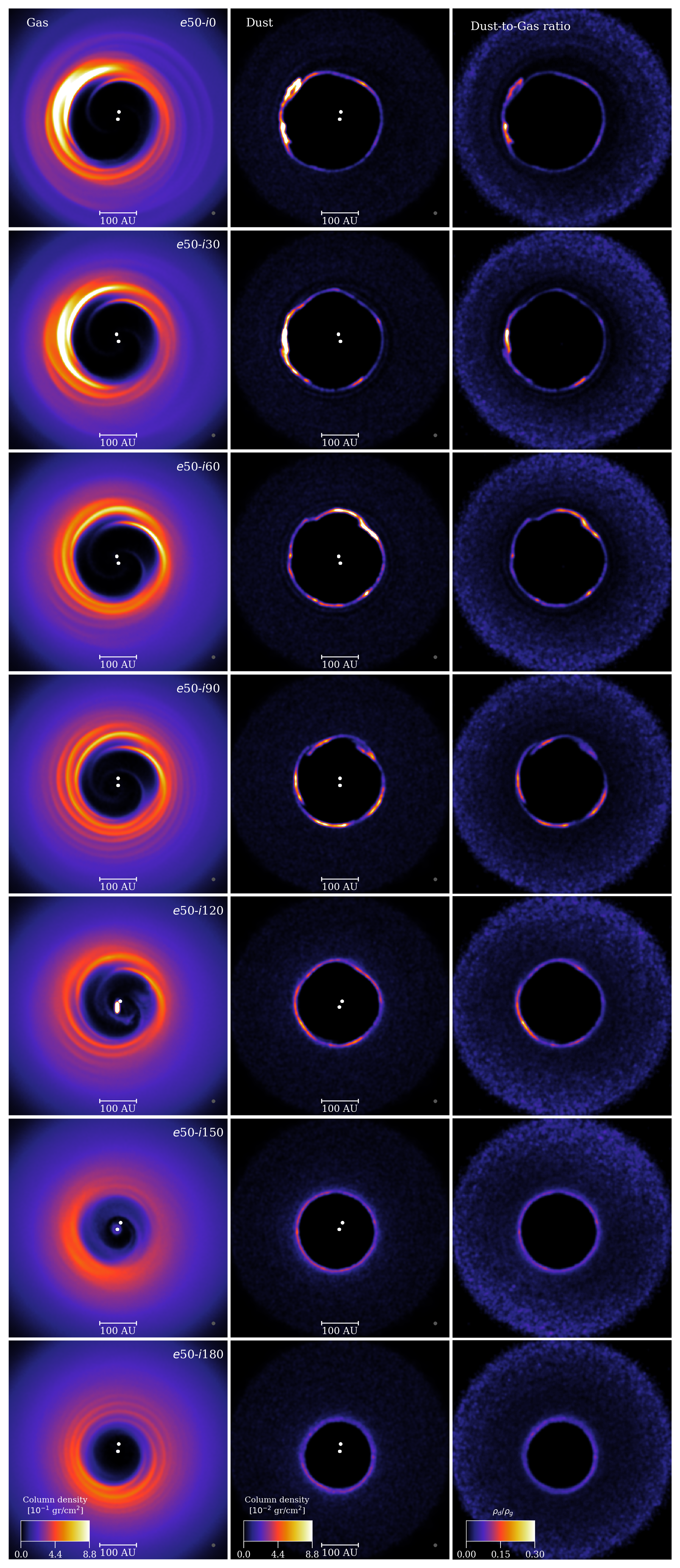}
\caption{Gas (first column), dust (middle column) surface density in $gr/cm^2$ and the dust-to-gas ratio in last column, after 50 binary orbits at $e_{\rm B}=0.50$. From upper to bottom are the different inclinations, $i_{\rm B} = \{0\degree,30\degree,60\degree,90\degree,120\degree,150\degree,180\degree\}$ respectively. The gray circle on the bottom-right corner of the two first columns represents the Gaussian kernel (5~au in diameter) used to smooth the images. This size is consistent with the highest ALMA angular resolution reached so far. We recall that in our simulations the physical quantities are computed using the smoothing length.}\label{fig:Density050}
\end{center}
\end{figure*}
\begin{figure*}
\begin{center}
\includegraphics[width=.55\textwidth]{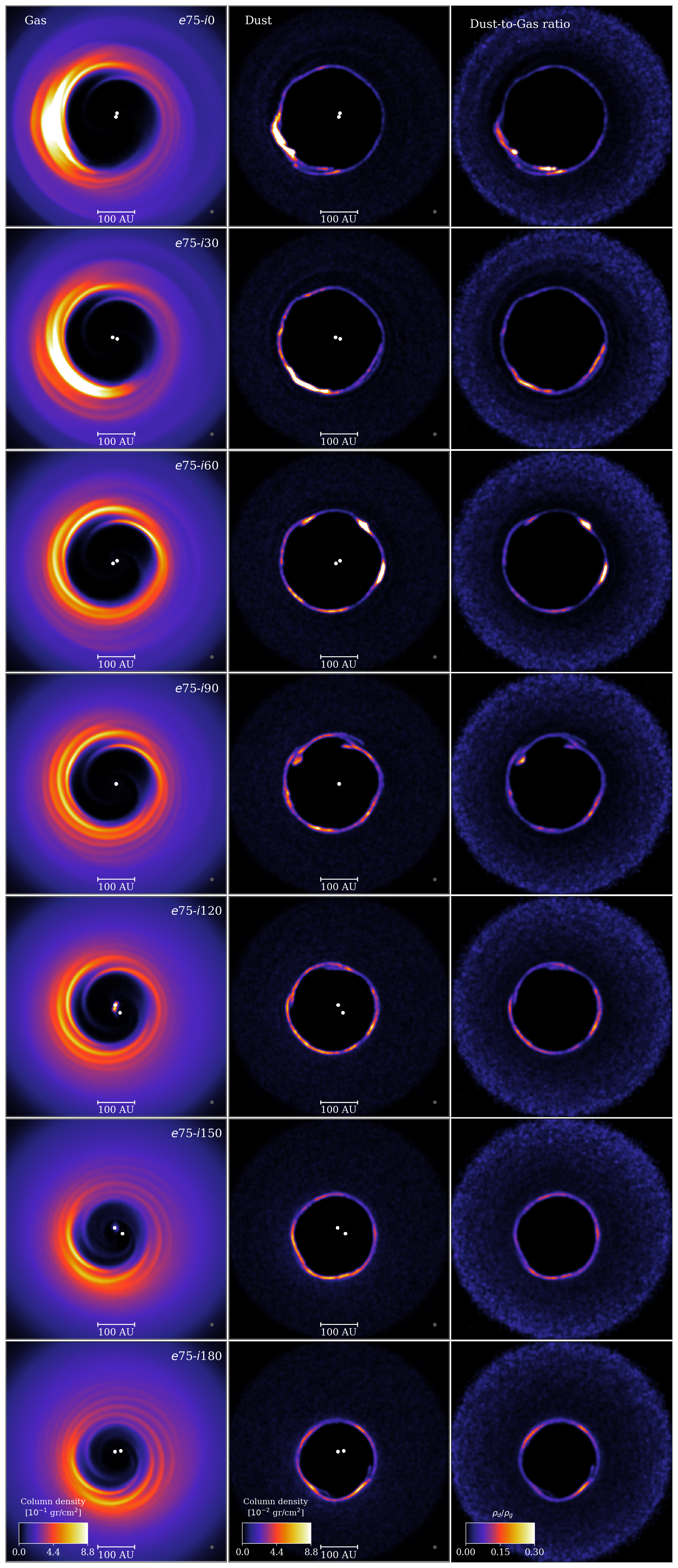}
\caption{Same as Fig.~\ref{fig:Density050}, but for the case $e_{\rm B}=0.75$.}\label{fig:Density075}
\end{center}
\end{figure*}

\subsection{Set of simulations}\label{sims}

The simulations are divided into two sets according to their eccentricity: $e_{\rm B}=0.50$ and $e_{\rm B}=0.75$. For each set, the inclinations are divided in prograde cases with $i_{\rm B}=\{0\degr,30\degr,60\degr\}$, the polar case with $i_{\rm B}=90\degr$, and the retrograde cases with $i_{\rm B}=\{120\degr,150\degr,180\degr\}$. The names of the simulations are listed in Table \ref{params}. In Figure \ref{orbits}, we show the binary eccentricity and orientation with respect to the circumbinary disc. In the following, the disc is always seen face-on with the binary inclined inside the inner cavity.

\section{Results}\label{Results}


Figures~\ref{fig:Density050} and \ref{fig:Density075} show the gas and dust surface density maps --- along with the dust-to-gas ratio --- for $e_{\rm B}=0.50$ and $e_{\rm B}=0.75$ (respectively). The CBD is shown after 50 binary orbits. At this evolutionary stage, the structures in the dust distribution have reached a quasi steady-state.

We observe features of different kind in the CBD. A {\em horseshoe}, defined as the main gas over-density at the edge of the disc cavity. A {\em dust ring} along the disc inner edge, caused by the radial drift of the dust particles. When the horseshoe traps most of the dust, a {\em large clump} of dust forms on top of it. We also observe that the dust can also get trapped in other regions along the dust ring. Remarkably, these {\em small clumps} are not bound to the horseshoe. In the following, we describe each of these disc features: as a function of $i_{\rm B}$ (Sect.~\ref{sec:proretro}) and $e_{\rm B}$ (Sect.~\ref{sec:ecc}).

\subsection{From prograde to retrograde cases}
\label{sec:proretro}

\begin{figure}
\begin{center}
\includegraphics[width=0.45\textwidth]{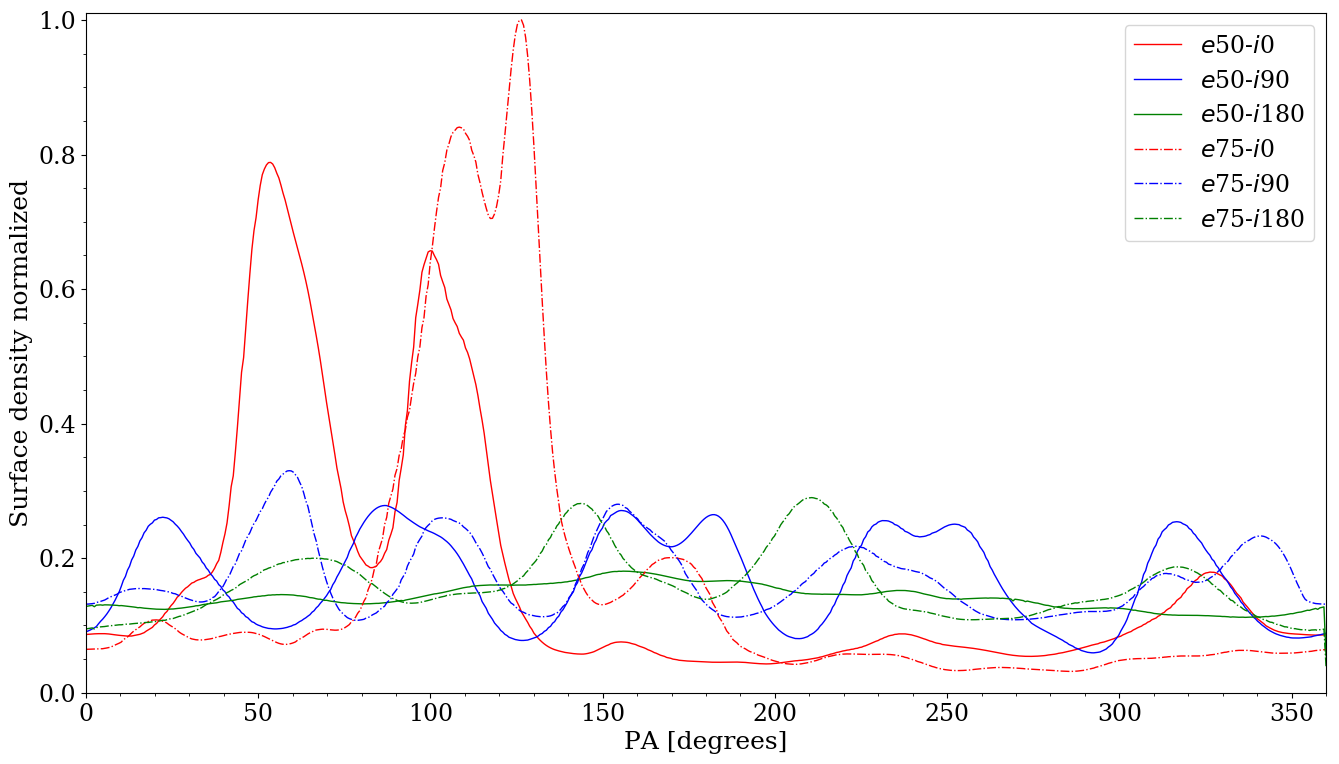}
\caption{Normalised dust surface density along the dust ring. Prograde ($i_{\rm B} = 0\degree$), polar ($i_{\rm B} = 90\degree$) and retrograde ($i_{\rm B} = 180\degree$) cases are shown in red, blue, and green (respectively). Solid and dashed lines correspond to $e_{\rm B} = 0.50$ and $e_{\rm B} = 0.75$, respectively.}\label{prof}
\end{center}
\end{figure}

\begin{description}

\item \emph{Prograde cases}: the most striking structure observed in Figs.~\ref{fig:Density050} and \ref{fig:Density075} is the horseshoe at the inner edge of the disc. This is seen only at inclinations $i_{\rm B}=\{0\degree,30\degree\}$, and for both eccentricities. This has already been reported in previous works of coplanar black hole binaries \citep{Shi12,DOrazio13,Farris14} and stellar binaries \citep{Ragusa17}. In particular, in the latter the authors report a dust concentration at the location of the horseshoe. This structure is comparable to the large dust clump observed in $e$50-$i$0, $e$50-$i$30, $e$75-$i$0, and $e$75-$i$30. However, we also observe the formation of small dust clumps outside the horseshoe. The differences between \textit{large} and \textit{small} clumps are their angular size and density. Quantitatively, the large clump is more than five times denser compared to the average dust density along the dusty ring; while the small clumps are only twice denser compared to the average value (see Figure \ref{prof}). In addition, the large clump tends to overlap with the horseshoe covering roughly $60\degree$ in azimuth; while small clumps cover smaller azimuthal sectors (less than $30\degree$). Remarkably, for $i_{\rm B}=60\degree$, both the horseshoe and the large clump disappear. Instead, several small clumps appear along the dust ring. When this happens, the binary-triggered spirals and streamers are the only gas structures observed.

The cases with $i_{\rm B}=\{0\degree,30\degree\}$ show a annular-shaped feature just outside the dense inner dust ring. This is easily seen in the dust-to-gas ratio maps of Figs.~\ref{fig:Density050} and \ref{fig:Density075}. This structure forms due to the action of the gaseous spiral arms on the dust, which modify the gas density in that region --- and hence the Stokes number. This speeds up the radial velocity of the dust particles, which eventually leads to the formation of a dusty gap in the disc.

\item \emph{Polar configuration}: for $i_{\rm B}=90\degree$, we see a set of $5$ small clumps evenly distributed along the dust ring (Fig.~\ref{prof}). The density of each of these clumps is about twice the average density of the dust ring.

\item \emph{Retrograde cases}: in this configuration, we observe a remarkable difference between $e_{\rm B}=0.5$ and $e_{\rm B}=0.75$, as explained in Section~\ref{sec:ecc}. Nevertheless, a common aspect is that the densest gas regions are displaced toward the companion's orbital apoastron. This is shown in the bottom rows of Figs.~\ref{fig:Density050} and \ref{fig:Density075}. The location of the densest region can be explained by the binary perturbations on the gas disc: the secondary star acts braking the surrounding gas. The deceleration is higher in the disc region closer to the companion \citep{Nixon11}.

\end{description}

Additionally, we note that the cavity size decreases with increasing binary inclination as found by \cite{Miranda15}. For instance, this effect can be easily seen by considering the shape of the dust ring. In addition, the cavity becomes eccentric and its centre does not match with the centre of mass of the system \citep{Dunhill15}. The coplanar cases ($i_{\rm B} = 0\degree$) present cavity sizes in agreement with the classic result of \citep{Artymowicz94} for different eccentricities.

We observe comparable values of the dust-to-gas ratio ($\epsilon=\rho_{\rm d}/\rho_{\rm g}$) in the large and small clumps. This is because for the large clump, the dust and gas densities are high; while for the small clump both densities are lower. The implications of the high dust-to-gas ratio values will be discussed in Section \ref{sec:dust-to-gas ratio}. 

Finally, we observe the formation of a circumprimary disc for $i_{\rm B}=120\degree$ and $i_{\rm B}=150\degree$. These kind of discs are likely transient and are hardly seen at this numerical resolution. This is because the low density of particles inside the cavity translates into a high numerical viscosity. Therefore, the circumstellar discs quickly drain into the stars, which are modelled as sink particles as in \cite{Price18}.

\subsection{Eccentricity 0.50 versus 0.75}\label{sec:ecc}

We find that the disc cavities are larger for $e_{\rm B}=0.75$ compared to $e_{\rm B}=0.5$, for the very same inclination. This is in agreement with \cite{Miranda15}. Also, the higher the binary eccentricity, the higher the density of the gas spirals and streamers. This is well seen for regrades cases with $e_{\rm B}=0.75$: the gas disc exhibits both prominent spiral structures and multiple spiral arms. The latter are concentrated in a specific azimuthal sector of the CBD. These disc features are not observed for $e_{\rm B}\leq0.50$, neither for coplanar configurations as in \cite{Ragusa17}. In Section~\ref{Clumps}, we show why the small dust clumps only form in retrograde configurations for $e_{\rm B}=0.75$.

\section{Discussion}\label{Discussion}

\subsection{Clumps}\label{Clumps}

As reported in Sect.~\ref{Results}, large and small clumps can form along the dust ring according to the binary parameters. The mechanism of dust trapping by a local azimuthal gas over-density --- namely the horseshoe --- has already been studied extensively in previous works \citep[e.g.,][]{Johansen04,Birnstiel13,Owen17,Ragusa17}. However, to the best of our knowledge, the formation of small dust clumps in CBDs has not been reported yet. These features are particularly prominent for $i_{\rm B} = 90\degree$, but they also appear for $i_{\rm B} = \{0\degree,30\degree\}$.

The main characteristic of the small clumps is that, although they form on top of local gas over-densities, they do not necessarily follow the gas.  This is in contrast to the large clump, explained above. Below, we focus in more detail on the formation of small clumps.

\subsubsection{Formation of a single small clump}\label{sec:formation}

For the coplanar retrograde ($i_{\rm B}=180\degree$) case, small clumps are only observed for $e_{\rm B} = 0.75$ (as opposed to $e_{\rm B} = 0.50$). This is because a higher eccentricity favours the formation of more prominent gas spirals and denser streamers. These gaseous features caused by the inner binary are crucial to trigger the small clump formation. Specifically, the binary-induced gaseous streams perturb the dusty ring through aerodynamical drag. The strength of the latter heavily depends on the Stokes number (see Eq.~\eqref{eq:Stokes}). For sake of simplicity, let's assume that the dust ring has a constant density, which is a reasonable approximation before any clump forms along the ring. Since the grain size is fixed, then the Stokes number only depends on the gas density.

\begin{figure}
\begin{center}
\includegraphics[width=0.5\textwidth]{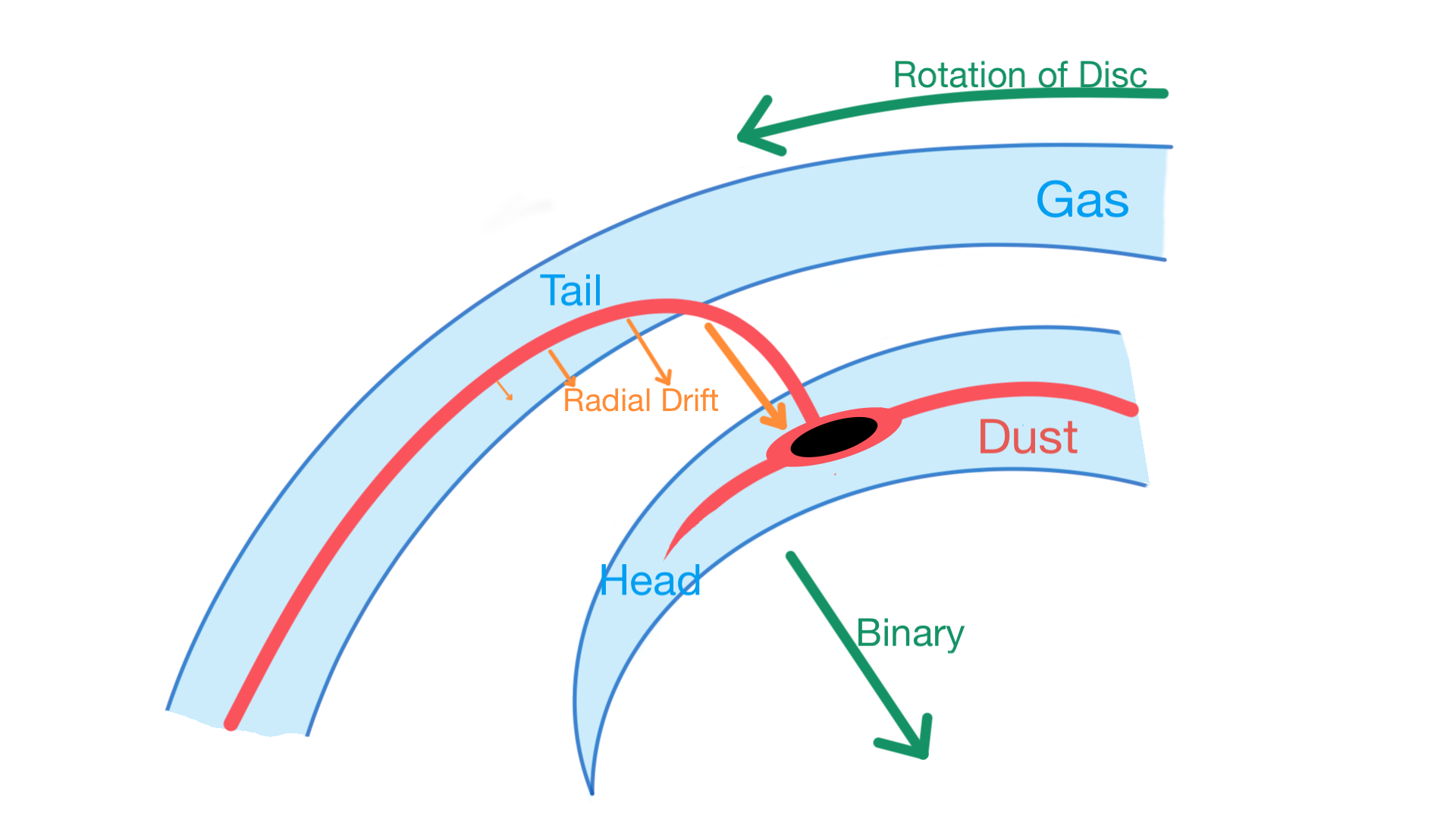}
\caption{Sketch of the mechanism of formation of a small clump. The gaseous spirals are represented in blue and the dust ring in red. The forming small clump is highlighted in black. The disc rotation and the binary centre of mass location are shown with green arrows. The length of the orange arrows indicates the magnitude of radial drift, which is higher in the region between the spirals.}\label{sketch1}
\end{center}
\end{figure}

The process of small clump formation is shown in Figure~\ref{sketch1} where we schematically represent the motion of the dust on top of the gaseous spirals. The inner spiral (called the \textit{head}) is caused by the secondary star, while the outer one (called the \textit{tail}) corresponds to the spiral formed in the previous orbit. The \textit{tail} is at a larger distance from the binary compared to the \textit{head}. The dust ring (in red) is deformed by the two gaseous spirals. The Stokes number in both spirals is less than one due to their high gas density, whereas it is higher in the region between the two spirals. For our disc parameters, 1~mm grains have $\rm St << 1$ in the spirals and $\rm St\sim1$ in between. Due to the strong coupling, the inner dust ring follows the \textit{head}, whereas the outer dust ring follows the \textit{tail}. In addition, the bending of the dust ring generates a significant radial density gradient. As a consequence, the dust particles in between the spirals radially drift towards the \textit{head}. This effect is the strongest for mm-sized grains because their Stokes number is close to one \citep{Weidenschilling77}. Therefore, millimetric dust is efficiently accumulated at the \textit{head} location, where a small clump begins to form.

\subsubsection{Evolution and behaviour of a single small clump}\label{evol}

Because of the periodic perturbation of the disc caused by the binary, gas spirals continuously form at a specific azimuthal sector of the disc. Therefore, after one binary orbit, the previous head becomes the tail (with a small clump attached to it) and the innermost gas stream becomes the head. This explains why dust clumps are periodically formed in the CBD in our models.

Once the clumps form, there are two possible dynamical outcomes: they can either be disrupted or keep growing. The survival and long-term behaviour of these \emph{individual structures} are key for grain growth, and consequently for planetesimal formation in the disc. The survival of the small clumps is related to the local Stokes number. 

The region where gaseous spirals are formed has high density -- clumps will have Stokes numbers less than one in there. Therefore, even though the small clump formation happens in that region, the clump can also be easily stretched and potentially disrupted. Such stretching is caused by the gradient of angular velocity within the clump due to interaction between the clump and the spiral. More specifically, the head of the spiral moves faster compared to the regions behind it. To characterise the evolution of the small clumps, we define their corresponding survival timescale as the time from their formation until their disruption. In all our simulations, we observe survival timescales of at least one orbital period at the clump radial distance. It is precisely when the clump completes the first orbit and returns to the spiral-forming region that it can be potentially disrupted. We also note that the inclination affects the clump survival time. For instance, in cases with $i_{\rm B} = 90\degree$, the binary torque does not strongly affect the azimuthal velocity gradient of the gas spiral. In this configuration, the small clumps survive for several tens of orbits.

\begin{figure}
\begin{center}
\includegraphics[width=0.5\textwidth]{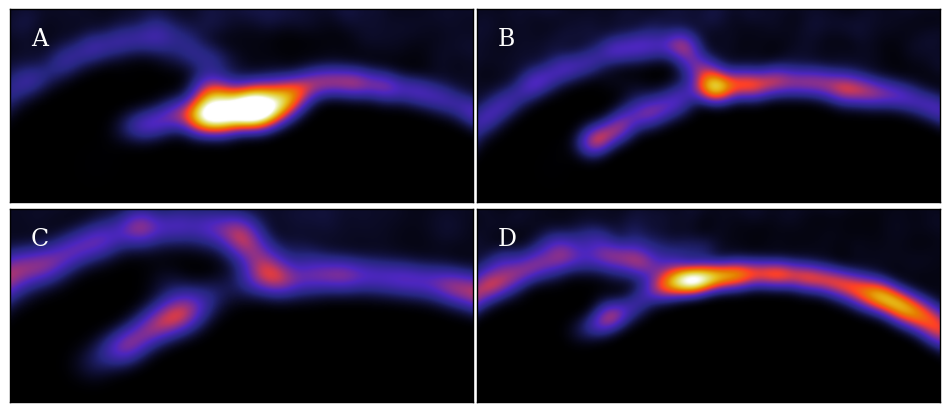}
\caption{The four possible scenarios that can experience a small clump after its formation. These examples are taken from the simulation $e$50-$i$90 at different times. In A, the clump is fed by a dust stream, without being disrupted. In B, we see a third clump forming in between a disrupted clump. In C and D, the dust streams feeds the back and front side (respectively) of a disrupted clump.}\label{sketch2}
\end{center}
\end{figure}

Besides disruption, the clump can also be fed after completing an orbit. Figure \ref{sketch2} shows all the possible scenarios that a small clump can experience. The first one, called A, happens when a dust stream falls exactly onto the small clump, making it grow. The other three scenarios (B, C, and D) lead to clump disruption. Once the small clump is disrupted, its remnants are later fed by the outer dust stream. However, it is worth noting that the previous clump never reforms as such. To sum up, it is possible to either generate a more massive clump (A); or to disrupt the main clump generating several ones (B, C, and D).

Throughout all the simulation, the circumbinary disc is periodically perturbed by the secondary star. This ensures the continuous formation of clumps as previously described. At the end of our simulations (i.e. after 100 binary orbits) the dust disc exhibits a similar morphology as the one observed after 50 binary orbits. Nevertheless, it is worth noting that the individual small clumps shown at 50 orbits are not necessarily the same as the ones present at the end of the simulation.

\subsubsection{Dust-to-gas ratio}\label{sec:dust-to-gas ratio}

Interestingly, if the dust-to-gas ratio becomes close to one then self-induced dust traps \citep{Gonzalez17} could appear in the disc. The increase of the dust-to-gas ratio could also potentially trigger the streaming instability \citep{Johansen07}. Therefore these could be \textit{sweet spots} for grain growth and planetesimal formation in the CBD. However, these dynamical effects are not seen in our simulations due to their short evolutionary time. In addition, SPH is not the most suitable method to capture streaming instability effects due to the two-fluid numerical scheme \citep{Laibe12a}. We did not run the models for longer, as the density approximation made in Sect.~\ref{sec:dust_pres}, namely $\rho = \rho_{\rm g}$, becomes less valid precisely as the dust-to-gas ratio increases.

\subsection{Clumps formation with different grain sizes}\label{dust_analysis}

So far, we have only discussed the large and small clump formation for one specific grain size, namely mm-sized particles. Here, we present simulations with  two other grain sizes: 100~$\mu$m and 1~cm. Figures \ref{e50_all_grains1} and \ref{e50_all_grains2} show the dust morphology for $e$50-$i$0 and $e$50-$i$90, respectively. In addition, we also show a test simulation without aerodynamic drag (labelled ``no drag"), where the dust particles behave as test particles.

Comparing Figures~\ref{e50_all_grains1} and \ref{e50_all_grains2}, we see that the grain size plays a crucial role in the formation of clumps. This is because the dust coupling (i.e. the Stokes number) depends linearly on the grain size, as shown in Eq.~\ref{eq:Stokes}. In particular, the proposed formation mechanism (see Sect.~\ref{sec:formation}) is the most efficient for dust grains with a Stokes number varying from one to slightly less than one along the orbit, which corresponds to $1$~mm for the disc parameters we have chosen. Nevertheless, for other disc parameters (e.g. disc mass, temperature and density profiles, etc.), the condition ${\rm St} \sim 1$ would correspond to a different grain size, which would form structures similar to the ones reported here.

Interestingly, the dust ring and the clumps are mainly caused by gas drag effects, and not only by the binary gravitational perturbations. For instance, in the no-drag simulations we do not observe any clumps or dusty rings along the cavity.

\subsection{Dust features as indicators of unseen stellar companion}\label{systems_known}

Direct observations of stellar companions in binary systems are particularly challenging. Specifically, there are strong limitations to properly resolve the separation between two stars. This is even worse if the companion is less massive and therefore fainter compared to the main star. However, here we have shown that some prominent disc features can be triggered by the gravitational interaction of a low-mass stellar companion inside the cavity. More specifically, the set of simulations of this work explore a modest but meaningful region of the vast space of parameters, namely the binary eccentricity and inclination. Hence, in principle, the disc features could be used to infer the orbit of a potentially unseen stellar companion.

\subsubsection{Remarkable dust structures in CBDs}
\label{sec:dust-structures}

\begin{description}
	\item \emph{Large dusty clump within a gas horseshoe.} These features appear in all our simulations with $i_{\rm B}\leq30\degree$. Both have the same properties as the ones reported by \citet{Ragusa17}. They are not produced by a vortex, and have a high contrast compared to the rest of the disc. Thus, a large dusty clump on top of a gas horseshoe could be an indicator of an inner companion with an orientation close to the disc plane.
	\item \emph{Embedded small clumps in a dusty ring.} For all our highly-inclined simulations ($60\degree \leq i_{\rm B}\leq120\degree$), we observed several small dusty clumps embedded in the disc. Interestingly, in the polar case, clumps are azimuthally equidistant between them. Therefore, several small clumps along the dust ring could indicate the presence of a \textit{highly inclined} inner companion. Note that the case $e$75-$i$180 shows a clump-ring structure too, therefore a highly eccentric and retrograde companion is also able to create the same feature. In Appendix~\ref{A1} we show that an inner planet-mass companion does not produce such structures, which allows us to set a lower mass threshold for structure formation in the CBD.
	\item \emph{Smooth dust ring.} All the cases that do not show any remarkable features (horseshoe or clumps) in the dust ring are included in this category. In the absence of structure it is hard to draw any conclusion on whether there is a single star or a binary system. Nevertheless, the inner cavity structure could help to infer the presence of an inner companion. For instance, during the early disc evolution, a large inner cavity of several tens of au strongly suggests the presence of a binary system inside the cavity. However, for more evolved discs and if no accretion is detected, the cavity is more likely to be caused by photoevaporative processes \citep{Alexander06,Owen16}
\end{description}

\begin{figure}
\begin{center}
\includegraphics[width=0.45\textwidth]{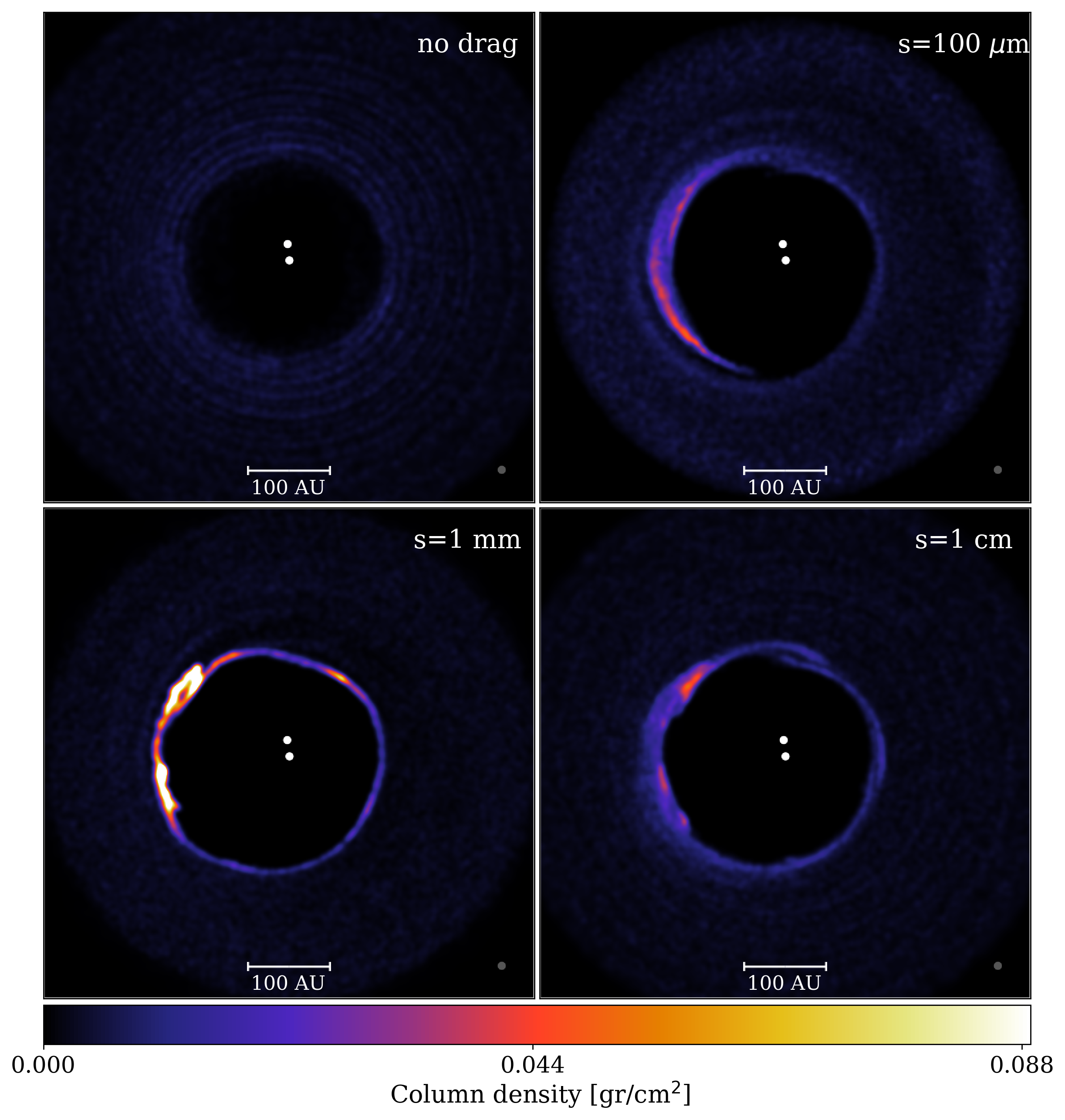}
\caption{Dust distribution of the case $e$50-$i$0 after 50 initial binary orbits, for three sizes of dust grains, plus one simulation without the aerodynamic drag. The dust grain sizes are 100 microns, 1 millimetre, and 1 centimetre. The no-drag simulation was made for $s=1$~mm.}\label{e50_all_grains1}
\end{center}
\end{figure}

\begin{figure}
\begin{center}
\includegraphics[width=0.45\textwidth]{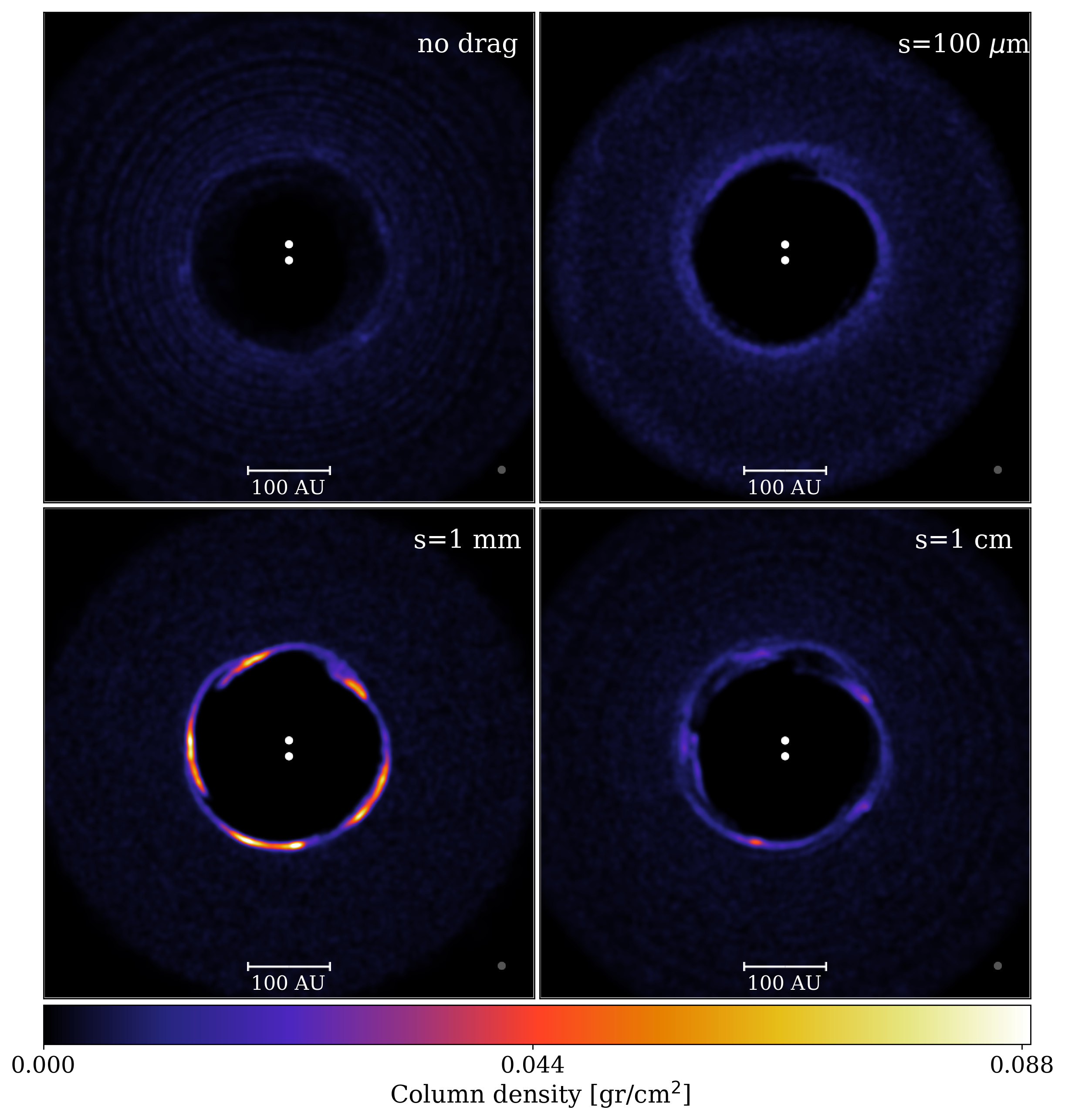}
\caption{Same as Fig.~\ref{e50_all_grains1}, but for the case $e$50-$i$90.}\label{e50_all_grains2}
\end{center}
\end{figure}

Caution is required when interpreting our results since this analysis mainly applies to grains with a Stokes number close to unity. See for instance the structures obtained for different grain sizes in Figures~\ref{e50_all_grains1} and \ref{e50_all_grains2}. HD~142527 is a notorious example where two different dust structures coexist: a large clump at millimetric wavelengths \citep{Boehler18} and small clumps at centimetric wavelengths \citep{Casassus15}.

When spirals are observed, their morphology and their azimuthal concentration in particular can provide further information on their dynamical origin. Besides an inner binary, flybys \citep{Cuello19b}, planets \citep{Dong15}, self-gravitating discs \citep{Dipierro15b,Forgan18}, or shadows \citep{Montesinos16,Montesinos18,Cuello19} can also produce spiral arms in the disc. The main difference is that the spirals caused by an eccentric inner companion are often multiple and well concentrated in one azimuthal sector of the disc, as opposed to the other mechanisms.

\subsubsection{The AB Aurigae system}\label{AB Aur}

AB Aurigae -- an Herbig Ae star of the A0 spectral type and mass $2.4 \pm 0.2\ M_{\odot}$ \citep{DeWarf03} -- exhibits a very complex morphology, both in gas and dust: i) multiple spiral arms in scattered light \citep{Fukagawa04,Corder05,Hashimoto11}, ii) a horseshoe-shaped dust trap \citep{Tang12,Pacheco16}, iii) and a large dust cavity at a distance from the star between $\sim 70$ and $100$~au \citep{Hashimoto11,Tang12}. A single planet has been proposed to explain some of the observed features \citep{Hashimoto11,Fuente17,Tang17}. \citet{Tang12} were only able to explain the cavity size by adding a body at $r\sim45$ au and $M=0.03M_{\odot}$; whereas \citet{Fuente17} managed to explain the emission of the dust disc by putting a Jupiter-mass planet at $r=94$~au. It is however challenging to explain all the aforementioned features simultaneously.

Instead, the binary scenario proposed by \citet{Price18} for HD~142527 seems more promising. As a matter of fact, there is a striking similarity between the structures observed in AB~Aurigae and those in HD~142527. \citet{Pirzkal97} gives an upper limit of 0.25 $M_{\odot}$ down to 60 au for an possible inner stellar companion in AB~Aurigae. It is worth noting that mass constraint at small radii is difficult to quantify, the mass upper limit could be greater. Therefore, a low mass ratio binary scenario for AB~Aurigae is reasonable. Based on the disc observations, our models suggest the presence of an inner stellar companion --- undetected so far.

\begin{figure*}
\begin{center}
\includegraphics[width=\textwidth]{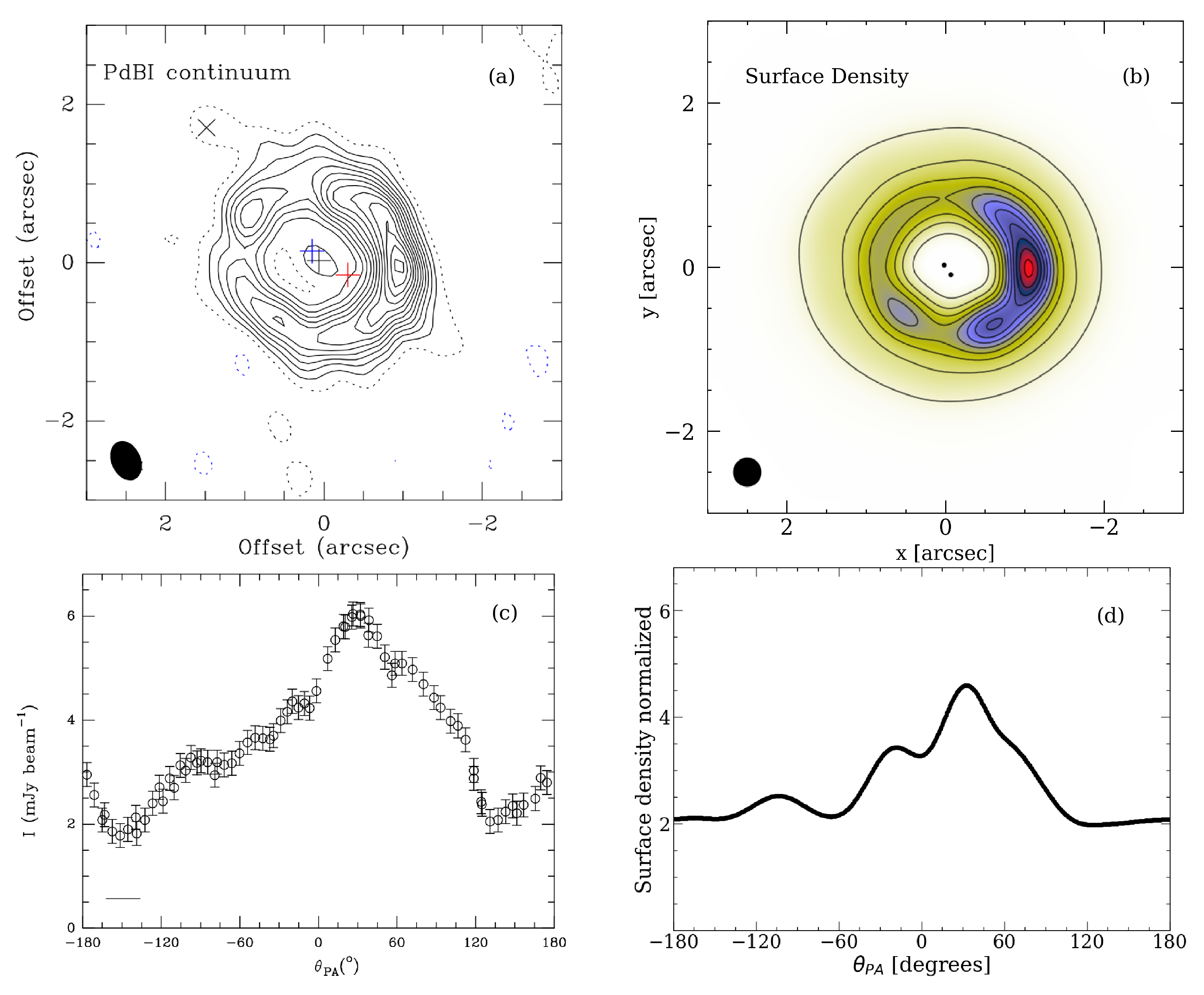}
\caption{Comparison between the observation at 1.3 mm of AB Aurigae (left column) and the dust distribution in $e$50-$i$60 after 100 orbits (right column). \textbf{(a)}: Dust continuum emission at 1.3~mm. The black cross represents the stellar peak. The blue and red crosses mark the peak of $^{12}$CO J=2$\rightarrow$1: highest blue-shifted and red-shifted peak respectively. \textbf{(b)}: Surface density map from $e$50-$i$60 convolved by a $50$~au beam (shown in the left corner), consistent with the observations in (a). \textbf{(c)}: 1.3~mm intensity along the dust ring. \textbf{(d)}: Surface density along the dust ring normalised to the average dusty ring density. The frame orientation is chosen so the dust over-density is roughly at the same azimuthal position as that in the the observation, and inclined $-23\degree$ with respect to the x-axis as the observed system \protect\cite{Tang12}. The $\theta_{\rm PA}$ in \textbf{(c)} and \textbf{(d)} represents the offset of the PA position at $121.3\degree$, measured from the north in a clockwise sense.  The observations in (a) and (c) are taken from Figures 1 and 11 in Tang, A\&A, 547, A84, 2012, reproduced with permission $\copyright$ ESO.}\label{comp}
\end{center}
\end{figure*}

As mentioned in Sect~\ref{sec:dust-structures}, the absence of a gas horseshoe and the multiple spiral arms suggest a high eccentricity and an inclination higher than $30\degree$. In particular, the case $e$50-$i$60 reproduces the observed dust distribution remarkably well (see Figure~\ref{comp}), simultaneously explaining the multiple and azimuthally concentrated gaseous spirals in the disc (not shown). Figure~\ref{comp} shows a comparison between the observed dust continuum emission at 1.3 mm \citep{Tang12} and the dust distribution in $e$50-$i$60, where the dust over-density is seen as a large clump due to beaming effects. In addition, \citet{Riviere-Marichalar19} very recently reported the observation of clumps in HCN -- a good tracer of cold and dense gas -- around the inner edge of the disc. This supports the idea that the dusty clumps might be real. It is worth to mention however that a bad coverage of the uv plane could produce artificial clumps in the reconstructed intensity map. Indeed, the 0.9 mm image presented by \citet{Tang17} shows a continuous inner ring, rather than clumps. Future observations at a higher angular resolution and with better uv plane coverage are required in order to reveal whether \emph{small clumps} are indeed embedded in the disc of AB~Aur.


In summary, our results strongly motivate the search for a stellar companion within the cavity of the disc around AB~Aurigae. More specifically, an i) unequal-mass, ii) eccentric, and iii) inclined stellar binary can potentially explain most (if not all) the observed disc features.



\section{Conclusions}\label{Conclusion}

We performed 3D SPH gas and dust simulations of circumbinary discs (CBDs) around  binaries with different eccentricities and inclinations. We considered unequal-mass binaries similar to HD~142527. This allowed us to characterise the disc morphology for different combinations of orbital parameters of the companion. The main conclusions of our work are the following:
\begin{enumerate}
\item An inner stellar companion with a low inclination ($i_{\rm B} \leq 30\degree$) with respect to the CBD is able to trigger a horseshoe-like structure in both the gas and the dust. Additionally, small dust clumps can also appear along the dusty ring. The latter had not been reported by previous works. 
\item For an inner stellar companion on a highly inclined orbit with respect to the CBD ($60\degree \leq i_{\rm B}\leq120\degree$) the dust ring breaks into small clumps, evenly distributed along the cavity.
\item For high eccentricities ($e_\mathrm{B} = 0.75$), the binary perturbations on the gas disc become stronger --- especially for retrograde cases ($120\degree \leq i_{\rm B}\leq180\degree$). This translates into denser gas structures: spirals, streams, and horseshoes. Therefore, the higher the eccentricity the easier the formation of clumps.
\item The small clump structure strongly depends on the Stokes number of the dusty ring. The formation mechanism is most efficient when the Stokes number is close to unity. A detailed description of its formation and evolution is given in Section~\ref{Clumps}.
\end{enumerate}

Circumbinary discs are often thought to be unfavourable systems for planetesimal formation, because of the high relative velocities expected among solid bodies \citep{Bromley15}. In this work, we have found that high dust-to-gas ratio clumps form in the inner regions of discs around unequal-mass, eccentric, and inclined binaries (especially for polar configurations). Such clumps could then constitute {\it sweet spots} for dust accumulation and grain growth \citep{Gonzalez17,Owen17}, suggesting that CBDs could potentially be efficient planetesimal cradles.

Within this context, polar circumbinary discs are of particular interest since we have shown they form stable and prominent dusty clumps ($e$50-$i$90 and $e$75-$i$90). Theoretical models first predicted the existence of this kind of polar discs \citep{Aly15,Martin17,Zanazzi18,Lubow18}, as the one very recently discovered in HD~98800 by \cite{Kennedy+2019}. It seems reasonable to think that planets will eventually form in these polar circumbinary discs. Based on that assumption, \cite{Cuello&Giuppone2019} showed that binaries with mild eccentricities are more likely to retain their circumbinary P-type polar planets (also known as {\it polar Tatooines}). In this regard, systems similar to $e$50-$i$90 are better candidates to host this type of polar planets.

Interestingly, the asymmetric disc features aforementioned can be much more easily detected than the binary itself. Considering the systems categorised as \emph{Giant Discs} by \citet{Garufi18}, several of them show multiple, asymmetric, relatively faint arm-like structures on large scales. Based on our results, we are inclined to think that there might be binaries in the cavities of several of those discs. In particular, we strongly suggest the presence of an eccentric and inclined inner companion in AB Aurigae. 

\section*{Acknowledgements}

We thank the anonymous referee for valuable comments and suggestions that have improved our work. Figures \ref{fig:Density050}, \ref{fig:Density075}, \ref{sketch2}, \ref{e50_all_grains1}, \ref{e50_all_grains2}, \ref{comp}(b), \ref{planet} and \ref{resolution} were made with {\sc splash} \citep{Price07}. The Geryon2 cluster housed at the Centro de Astro-Ingenieria UC was used for the calculations performed in this paper. The BASAL PFB-06 CATA, Anillo ACT-86, FONDEQUIP AIC- 57, and QUIMAL 130008 provided funding for several improvements to the Geryon/Geryon2 cluster. The authors acknowledge support from CONICYT project Basal AFB-170002. PPP and JC acknowledge support from Iniciativa Cient\'ifica Milenio via the N\'ucleo Milenio de Formaci\'on Planetaria. NC acknowledges financial support provided by FONDECYT grant 3170680. This project has received funding from the European Union's Horizon 2020 research and innovation programme under the Marie Sk\l{}odowska-Curie grant agreement No 823823.

\bibliographystyle{mnras}
\bibliography{Biblio}

\appendix

\section{Planetary companion in the polar case}\label{A1}

\begin{figure}
\begin{center}
\includegraphics[width=0.5\textwidth]{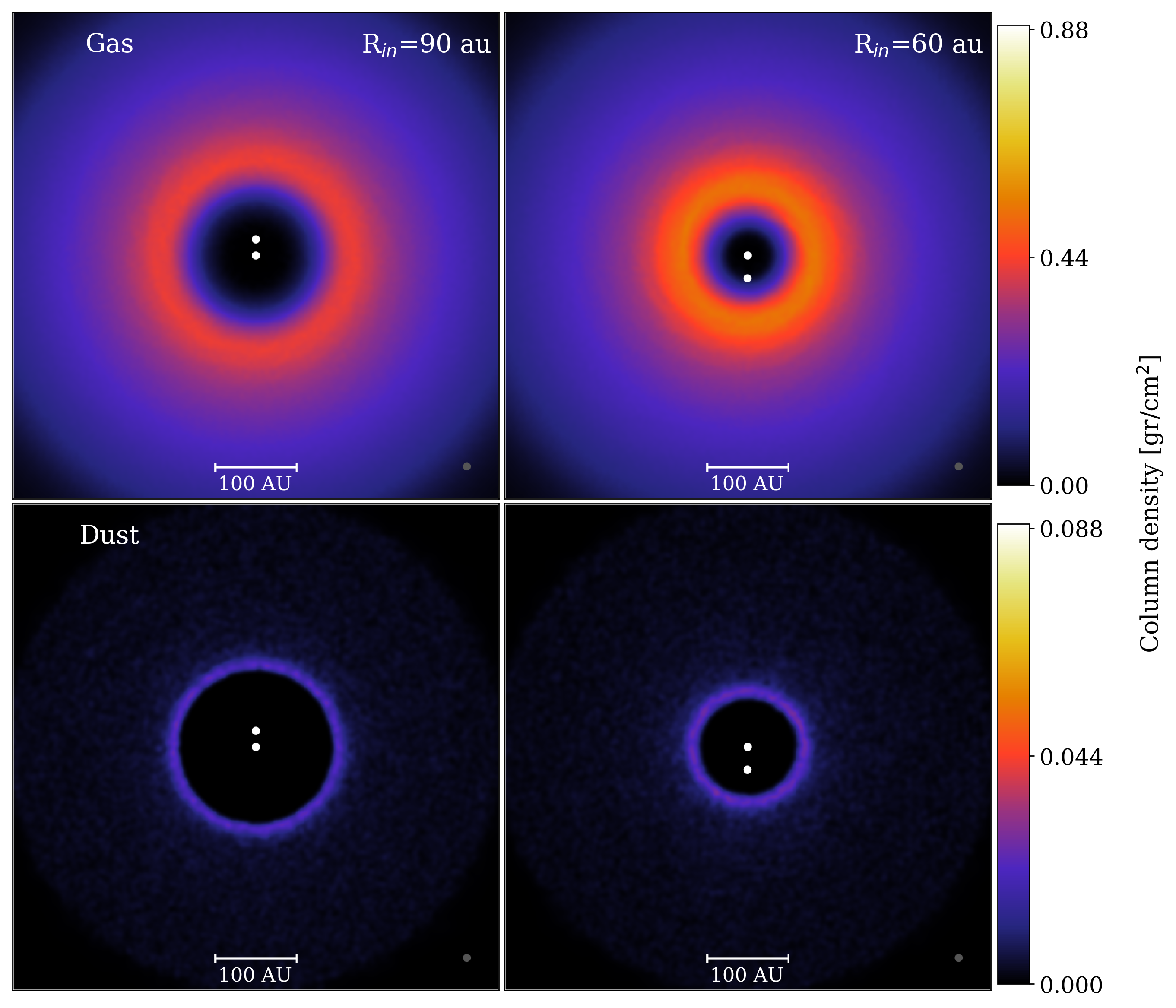}
\caption{Gas (top panels) and dust (bottom panels) morphology for the case $e50$-$i90$ after 50 binary orbits, but with a companion mass reduced to 10~$M_\mathrm{J}$ (i.e. 50 times less massive). The initial disc inner edge is set at 90~au and 60~au in the left and right panels, respectively. A planetary companion is not able to trigger the formation of dusty clumps in the disc.}\label{planet}
\end{center}
\end{figure}

To test whether the structures we found in the dust disc can also be triggered by planetary-mass companions, we performed two simulations with the very same setup as $e$50-$i$90, but with a companion mass of 10~$M_\mathrm{J}$. In addition, we considered different values for the initial disc inner edge: $R_{\rm in}=90$ au and $R_{\rm in}=60$ au. Given the reduced strength of the gravitational perturbations, the reduced cavity size allows us to bring material closer to the inner planetary companion. In Figure \ref{planet}, we show the dust and gas distributions after 50 planetary orbits. Regardless of the value of $R_{\rm in}$,  the structures are different from the ones obtained for a stellar companion (see middle row in Fig.~\ref{fig:Density050}). The only remarkable feature is the smooth dust ring. No small clumps nor spirals are observed. In summary, the latter features can only be triggered by a stellar companion for the configuration considered.

\section{Numerical tests for small clump formation}
\label{A2}

\begin{figure*}
\begin{center}
\includegraphics[width=\textwidth]{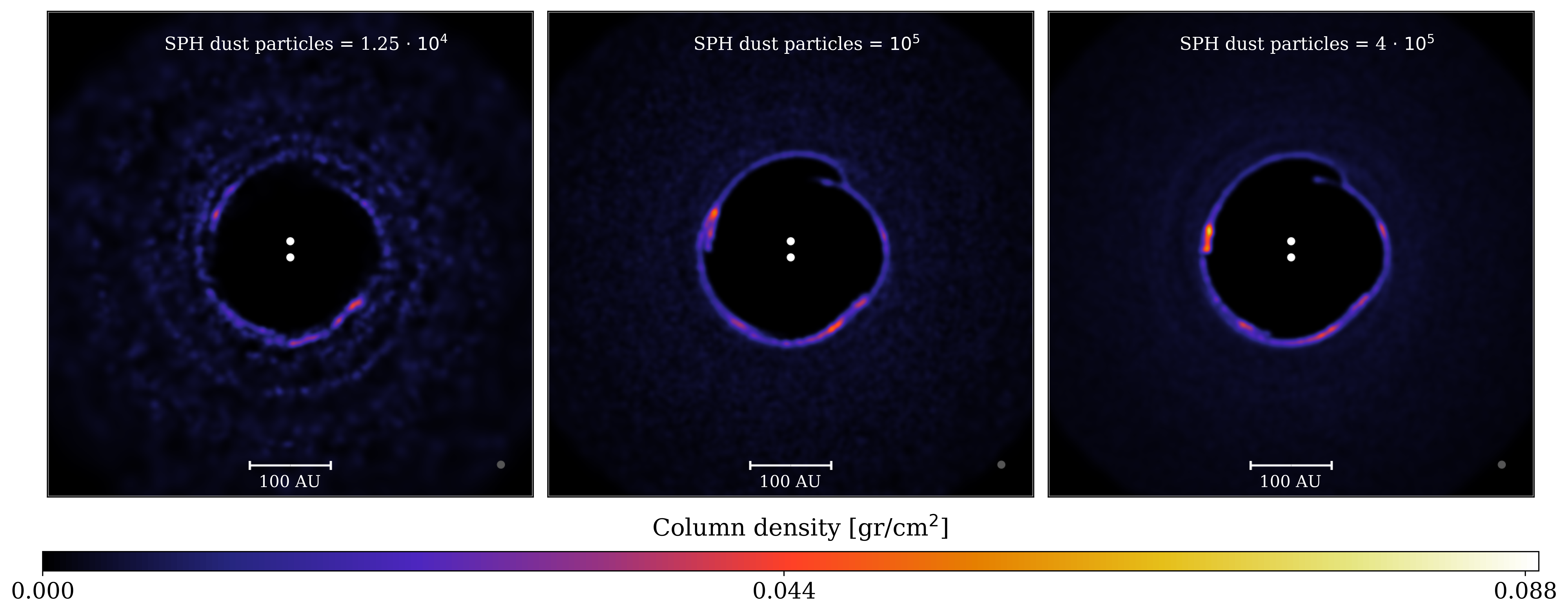}
\caption{Dust distribution for the case $e$50-$i$90 after 11 binary orbital periods for different resolutions in dust: $1.25\cdot10^4$ (left), $10^5$ (middle), and $4\cdot10^5$ (right) SPH dust particles. The gas resolution is fixed to $10^6$ SPH gas particles for all the simulations. The formation of dusty clumps is not affected by the dust resolution.}\label{resolution}
\end{center}
\end{figure*}

In order to test whether or not the small dusty clumps reported in this work were caused by numerical effects, we performed a convergence test for different resolutions in dust. This was done for the case $e$50-$i$90 (i.e. our more representative example) at three resolutions: $1.25\cdot10^4$, $10^5$ (this work), and $4\cdot10^5$ dust particles; keeping the gas resolution fixed to $10^6$ particles. These three simulations are shown in the left, middle, and right panels of Figure~\ref{resolution} (respectively). We observe that the low and high resolution tests exhibit the same structures (dust ring plus small clumps) as the simulation with $10^5$ dust particles. Their azimuthal positions are identical for the resolutions considered. We also see that the proposed mechanism to form small clumps by bending the dust ring (see Sect.~\ref{sec:formation}) still holds --- regardless of the number of dust particles. The five clumps seen in Fig.~\ref{fig:Density050} do not appear here because of the earlier evolutionary stage of the disc (11 binary orbits instead of 50). Since we observe emergent small clumps, it is reasonable to expect these features to appear eventually. Based on these results, we thus conclude that the formation of small dusty clumps is a physical process, which is properly captured at the resolution of $10^5$ dust particles.

\begin{figure}
\begin{center}
\includegraphics[width=0.4\textwidth]{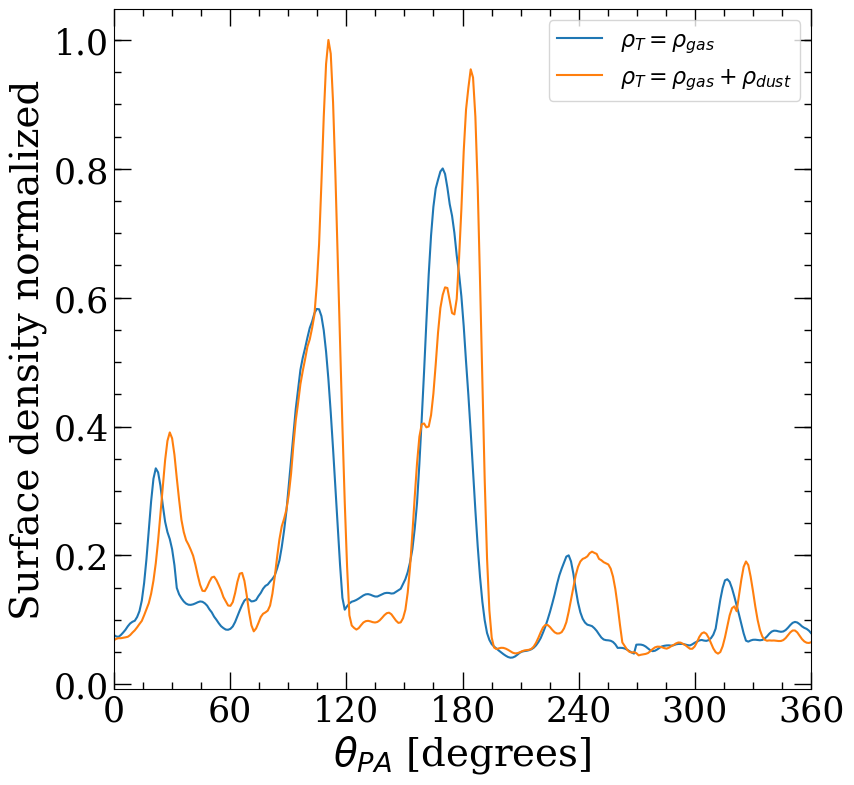}
\caption{Azimuthal density profile along the dust ring after 21 binary orbits. The blue line corresponds to $e$50-$i$60 with the Stokes number computed with the approximation ($\mathrm{St}\propto \rho_{\mathrm{g}}^{-1}$). Instead, the orange line corresponds to $e$50-$i$60 without the approximation ($\mathrm{St}\propto (\rho_{\mathrm{g}}+\rho_{\mathrm{d}})^{-1}$, where $\rho_{g}$ and $\rho_{d}$ are the gas and dust density respectively).}\label{approx}
\end{center}
\end{figure}

Finally, in order to test whether the formation of small dusty clumps is affected by our approximation of the Stokes number (see Eq. \ref{eq:Stokes}) we performed a shorter simulation without the approximation, i.e., computing $\mathrm{St} \propto \rho_T^{-1} = (\rho_{\mathrm{g}} + \rho_{\mathrm{d}})^{-1}$. We did this for $e$50-$i$60 because it is the case that shows the highest dust-to-gas ratio values. Hence, it presents the most significant difference between the Stokes number computed with and without the approximation. Figure~\ref{approx} shows the azimuthal density profile of both models. We see that regardless of the way we compute the Stokes number, the small clumps form, and also do it at approximately the same azimuthal position. We therefore conclude that the approximation does not affect our results in any significant way.

\section{Disc vertical scale-height at different radial distances}\label{A3}

\begin{figure}
\begin{center}
\includegraphics[width=0.5\textwidth]{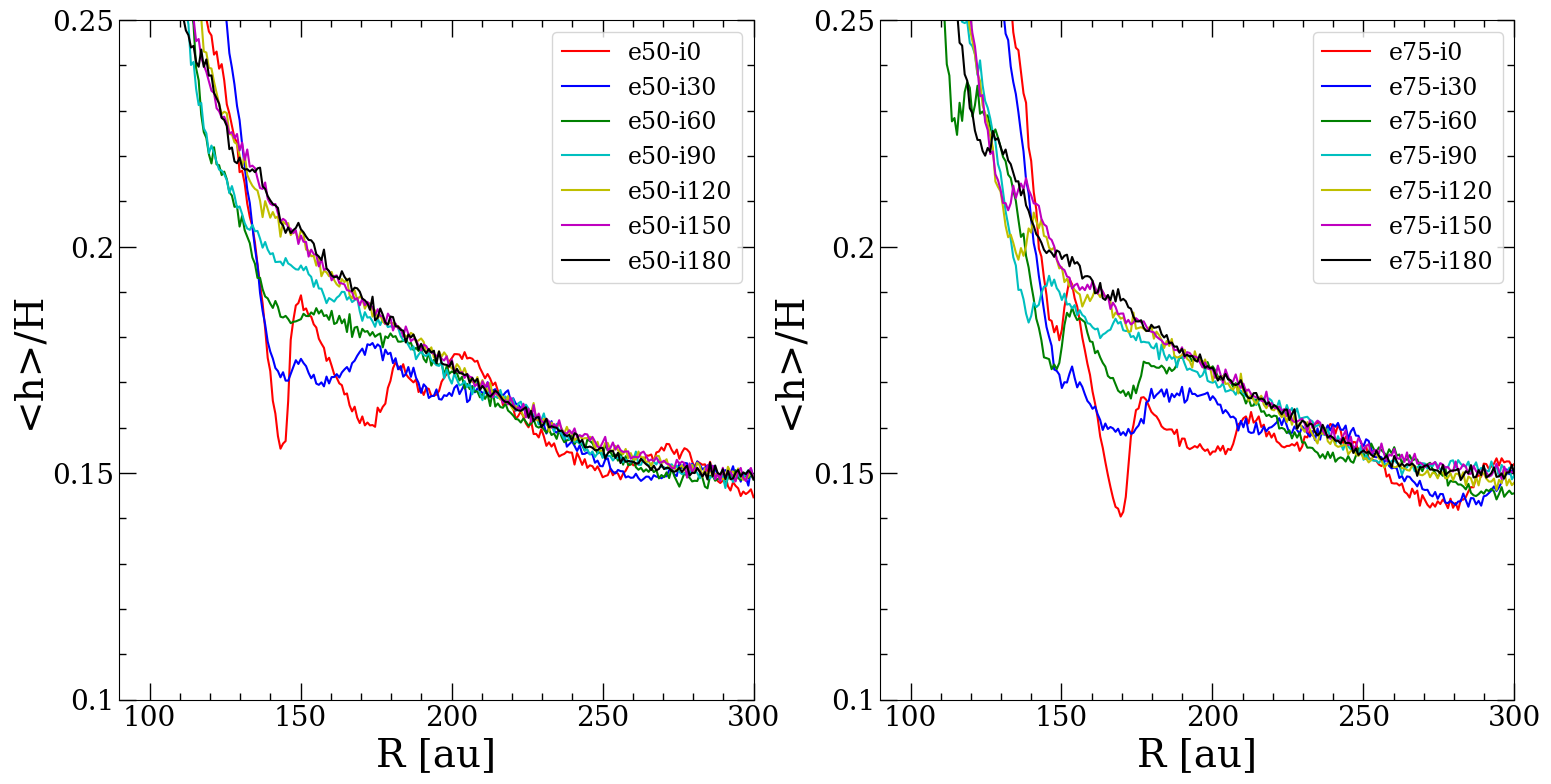}
\caption{Radial profiles of $<h>/H$ after 50 binary orbits for all the simulations with $e_{\rm B}$ = 0.50 (left panel) and with $e_{\rm B}$ = 0.75 (right panel).}\label{hH}
\end{center}
\end{figure}

In Figure \ref{hH} we show $<h>/H$ for our simulations with $e_{\rm B}$ = 0.5 (left panel) and with $e_{\rm B}$ = 0.75 (right panel). Since $<h>/H<1$, the disc is properly resolved in the vertical direction. The value of the Shakura--Sunyaev viscosity $\alpha_{\rm SS}$ can be easily inferred from the values of $<h>/H$ in Figure \ref{hH}. In this case, it is of the order of $0.005$ as mentioned in the text and as in the simulations in \citet{Price18}.

\end{document}